# Plasmonic Bound States in the Continuum to Tailor Light-Matter Coupling

Andreas Aigner[1,*], Andreas Tittl[1], Juan Wang[1], Thomas Weber[1], Yuri Kivshar[2], Stefan A. Maier[1,3,4,*], and Haoran Ren[3,*]

1) Chair in Hybrid Nanosystems, Nano-Institute Munich, Faculty of Physics, Ludwig-Maximilians-University Munich, Munich, 80539, Germany.
2) Nonlinear Physics Center, Research School of Physics Australian National University, Canberra, ACT 2601, Australia.
3) School of Physics and Astronomy, Monash University, Clayton, Victoria 3800, Australia.
4) Department of Physics, Imperial College London, London SW7 2AZ, United Kingdom.

*E-mails: andreas.aigner@physik.uni-muenchen.de; stefan.maier@monash.edu; haoran.ren@monash.edu

**ABSTRACT**

Plasmon resonances play a pivotal role in enhancing light-matter interactions in nanophotonics, but their low-quality factors have hindered applications demanding high spectral selectivity. Even though symmetry-protected bound states in the continuum with high-quality factors have been realized in dielectric metasurfaces, impinging light is not efficiently coupled to the resonant metasurfaces and is lost in the form of reflection due to low intrinsic losses. Here, we demonstrate a novel design and 3D laser nanoprinting of plasmonic nanofin metasurfaces, which support symmetry-protected bound states in the continuum up to $4^{th}$ order. By breaking the nanofins' out-of-plane symmetry in parameter space, we achieve high-quality factor (up to 180) modes under normal incidence. We reveal that the out-of-plane symmetry breaking can be fine-tuned by the triangle angle of the 3D nanofin meta-atoms, opening a pathway to precisely control the ratio of radiative to intrinsic losses. This enables access to the under-, critical-, and over-coupled regimes, which we exploit for pixelated molecular sensing. Depending on the coupling regime we observe negative, no, or positive modulation induced by the analyte, unveiling the undeniable importance of tailoring light-matter interaction. Our demonstration provides a novel metasurface platform for enhanced light-matter interaction with a wide range of applications in optical sensing, energy conversion, nonlinear photonics, surface-enhanced spectroscopy, and quantum optics.

**INTRODUCTION**

Controlling and enhancing light-matter interactions is the foundation of nanophotonics. Over the last decades, surface plasmons have attracted much attention due to their subwavelength confinement of light, which can be exploited to remarkably enhance light-matter interactions. Plasmon resonances in the form of localized and propagating modes at metal-dielectric interfaces have been extensively used for plasmon-enhanced sensing[1,2], fluorescence[3,4] and Raman spectroscopy[5,6], photoacoustics[7,8], photocatalysis[9,10], photovoltaics[11,12], nonlinear optics[13,14], to name a few. However, these plasmon resonances typically feature low quality factors (Q-factors) of around 10 due to the intrinsic loss in metals[15], which has limited their practical use, as many applications demand high spectral selectivity. Even though arrays of metallic nanostructures may support narrow surface lattice resonances (SLRs) with roughly 10-fold Q-factor improvement[15], they are extremely sensitive to both structural disorders and illumination conditions. Furthermore, they typically demand very large structured areas as well as embedment in a homogenous medium[16,17,18], hindering practical applications.

Meanwhile, bound states in the continuum (BICs) were introduced to nanophotonics. Originating from quantum mechanics[19,20], BICs are modes with an infinite Q-factor in an open system which cannot couple to any radiation channel propagating outside the system[21,22,23]. Small perturbations can destroy pure BICs and result in quasi-BIC modes with finite Q-factors which become accessible to the far field. Among different types of BICs, symmetry-protected BICs have gained by far the most attention. To date, they have been mostly realized in low-loss dielectric metasurfaces using in-plane symmetry-broken geometries[24,25,26,27]. For many photonic applications such as optical sensing[28], energy conversion[29], and light emission[30], not only high Q-factors but also strong near-field enhancement accompanied by a high coupling efficiency is essential. The latter is critical for maximizing light-matter interactions but cannot be obtained by lossless dielectric metasurfaces due to their mismatch between the radiative and intrinsic losses[31]. Most of the impinging energy is not coupled to the resonant metasurfaces and hence lost in the form of reflection[32,33]. Although symmetry-protected BICs were recently observed in plasmonic metasurfaces[34,35,36,37], they rely on symmetry breaking in momentum space via oblique incident angles, rendering them not suitable for most applications. Therefore, it remains a great challenge to achieve a metasurface platform simultaneously supporting tunable high Q-factors at normal incidence as well as high light-matter coupling.

Here, we demonstrate the design and 3D laser nanoprinting of a plasmonic nanofin metasurface (PNM) supporting multiple out-of-plane symmetry-protected BICs, which we employ for pixelated molecular sensing in different coupling regimes. The PNM consists of an array of identical 3D triangular nanofin building blocks fabricated in a polymer matrix and coated with gold (Fig.1a). By breaking the out-of-plane symmetry of the nanofins through the triangle angle $\alpha$ (defined in Fig.1b), the nanofin transforms from a rod-like to a triangular structure supporting symmetry-protected quasi-BICs. The 1$^{st}$ and 2$^{nd}$ order modes show Q-factors of up to 105 and 180, respectively. We show that the excitation efficiency (represented by the absorption) and Q-factor of the PNM can be flexibly tailored through $\alpha$ in parameter space (Fig.1c). Precise engineering of the ratio of radiative to intrinsic losses allows tuning of the PNM from the under- (UC) to critical- (CC) to the over-coupling (OC) regime. Consequently, we constructed a coupling-tailored PNM array that covers all coupling regimes to perform pixelated molecular sensing (Fig.1d). The necessary broad wavelength coverage in the mid-infrared (MIR) wavelength region was achieved by scaling the size of PNMs in small steps. We demonstrate that molecular analytes can introduce negative, positive, or no modulations to the quasi-BICs depending on the coupling efficiency between light and the used PNM pixels (Fig.1e). In addition to high Q-factors and field enhancements, our results suggest that for maximal sensitivity the radiative decay rate of the resonant metasurface needs to be tailored as well.

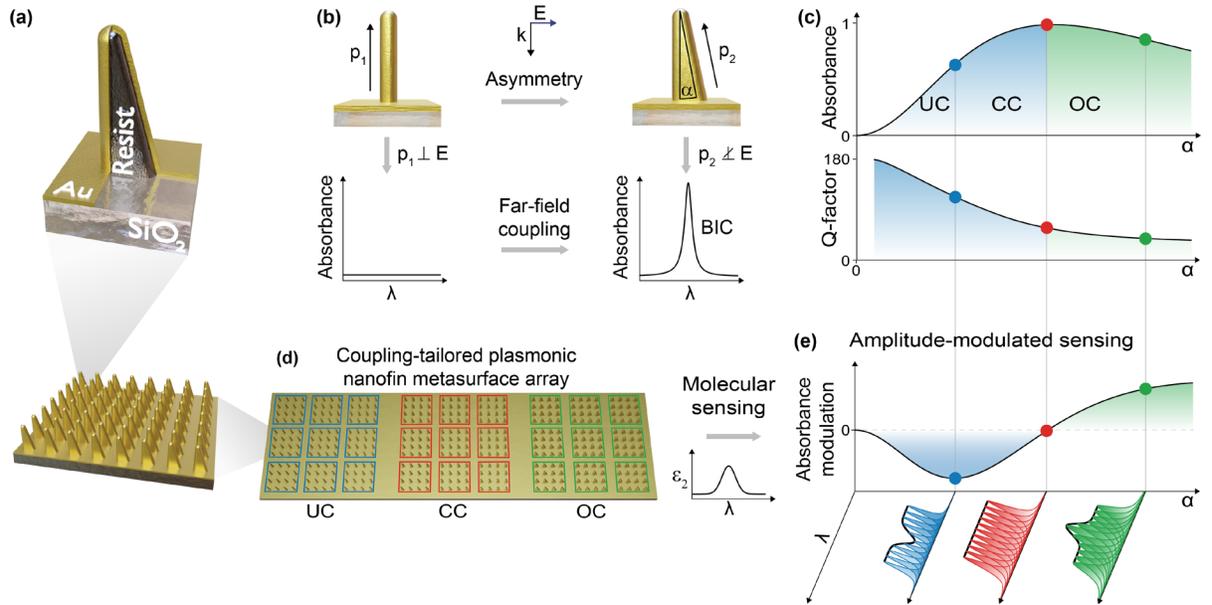

**Fig.1: Principle of PNMs and their coupling-tailored sensing capabilities. (a)** Schematic illustration of a PNM consisting of standing polymer triangles covered in an optically dense layer of gold on a SiO$_2$ substrate. **(b)** A symmetric, standing dipolar rod featuring a dipole moment $p_1$ normal to the incoming electric field $E$, thus no coupling is possible. Transforming the rod into a triangle allows coupling of its dipole moment $p_2$ to $E$, controlled by the asymmetry parameter α. **(c)** Tuning α allows for precise control over the radiative loss of the PNMs, capable of tailoring the absorbance from zero to unity. The under-coupled (UC), critically-coupled (CC), and over-coupled (OC) regimes are marked with different colors. **(d)** Schematic illustration of a coupling-tailored PNM array for molecular sensing, which consists of size-scaled PNM units in different coupling conditions. **(e)** Absorbance-modulated sensing based on the coupling-tailored PNM array, in which the analyte's signal appears as envelope of the scaled PNM resonances. The sensitivity strongly depends on the coupling regime, with the lowest sensitivity around the CC condition and the highest in the UC and OC regime with negative and positive modulations, respectively.

## RESULTS

### Design of BIC-enhanced plasmonic metasurfaces

All investigated structures are based on arrays of vertically oriented, gold coated triangles defined by their height $h$, diameter $d$, triangle angle $\alpha$, and periodicity $\Lambda$ (Fig.S1). The gold layer is optically opaque (with thickness $> 40\mathrm{nm}$), yielding zero transmittance and a simplified relation between absorbance $A$ and reflectance $R$ as $A = 1 - R$. A numerical analysis of different coating thicknesses reveals a minimal required gold thickness of $30\,\text{nm}$ in the MIR (Fig.S2). To investigate the photonic behavior of the PNM, we numerically studied the out-of-plane symmetry breaking in both momentum (k-) and parameter space by tilting the incident light ($\theta = 0 - 25°$) on a symmetric rod structure and breaking the structural asymmetry ($\alpha = 0 - 25°$) under normal incidence, respectively. The MIR absorption spectra of a PNM with $h = \Lambda = 3.5\,\mu\text{m}$ and $d = 0.7\,\mu\text{m}$ reveal three high Q-factor modes (Fig.2a). At the Γ-point, where both excitation and structure are symmetric ($\theta = \alpha = 0°$), the two modes around $6 - 8.5\,\mu\text{m}$ and $5\,\mu\text{m}$ vanish and the PNM acts like a gold mirror with unity reflectance. This indicates the existence of two symmetry-protected BICs at $8.3\,\mu\text{m}$ (BIC1) and $4.8\,\mu\text{m}$ (BIC2) which can be turned into leaky quasi-BICs by breaking their inversion symmetry either in momentum or parameter space. We attributed the third mode around $3 - 4\,\mu\text{m}$ to an in-plane SLR that does not vanish at the Γ-point but features an accidental BIC at $\alpha = 18°$ and a Q-factor of up to $1300$ (Fig.S3). As our experiments discussed below show, the SLR is drastically quenched due to its instability regarding fabrication and measurement imperfections.

To elaborate more on the physical origin of the different resonances, we inspected their mode profiles at the points of maximum absorption in parameter space marked by red lines in Fig.2a. The corresponding electric and magnetic field distributions of BIC1, BIC2, and the SLR are shown in Figs.2b, 2c, and 2d, respectively. The two symmetry-protected quasi-BIC modes clearly exhibit out-of-plane dipolar profiles of the 1st (BIC1) and 2nd (BIC2) order, showing electric field vectors perpendicular to the excitation and circulating magnetic field vectors around the nanofins (Figs.2b and 2c). Intuitively, these out-of-plane dipoles cannot couple to incoming light at the Γ-point with the scalar product of the nanofins' dipole moment and the impinging electric field being zero, as illustrated in Fig.1b. Note that the gold layer on top of the substrate plays a crucial role by reducing the radiative decay channels to one (reflection). In contrast to the out-of-plane BIC modes, the mode profile of the SLR in Fig.2d resembles an in-plane electric dipole with strong electric fields between neighboring nanofins and an out-of-plane circulating magnetic field around them. To study the effect of asymmetry on the light-matter coupling, Fig.2e shows the absorbance maximum, the Q-factor, and the surface field enhancement of BIC1 for the asymmetry parameters $\theta$ and $\alpha$, revealing an almost mirror symmetric behavior for the two symmetry breaking strategies. By adjusting $\theta$ or $\alpha$, these mode properties can be easily controlled resulting in maximal Q-factors in parameter space of 105 for BIC1 (blue curve in Fig.2e) and 180 for BIC2 (Fig.S4) as well as a maximum surface field enhancement of 750 and $1.98 \times 10^3$, respectively. The latter values and their strong dependency on the assumed nanofin tip shape is further discussed in Supplementary Note 1. In the following, we focus on symmetry breaking in parameter space, which allows for direct multiplexing of PNMs with different asymmetries on the same device without changing the measurement conditions, making it more desirable for practical applications.

For a resonant cavity, the Q-factor is typically comprised of two parts[31], the radiative $Q_{\text{rad}}$ and the intrinsic $Q_{\text{int}}$, which are associated with the radiative $\gamma_{\text{rad}}$ and intrinsic $\gamma_{\text{int}}$ losses, respectively. The Q-factor can be defined as: $Q = \left(\frac{1}{Q_{\text{rad}}} + \frac{1}{Q_{\text{int}}}\right)^{-1} = \frac{\omega_0}{2(\gamma_{\text{rad}}+\gamma_{\text{int}})}$, where $\omega_0$ is the angular frequency of the resonance. It is common for dielectric metasurfaces (with a negligible $\gamma_{\text{int}} \approx 0$) to have a diverging Q-factor around the BIC condition, namely $Q \approx Q_{\text{rad}} = \frac{\omega_0}{\gamma_{\text{rad}}} \to \infty$ for $\gamma_{\text{rad}} \to 0$. This leads to the for BICs typical relation between the Q-factor and a structural asymmetry factor $AF$: $Q^{-1} \propto AF^2$ [27]. In contrast to dielectric systems, we must separate the radiative and intrinsic losses in our PNMs for which we employ temporal coupled-mode theory (TCMT). We model our PNM with a single resonator coupled to one radiative and one intrinsic channel (Fig.2f) as described in Supplementary Note 2 in more detail. Accordingly, the absorbance $A$ of the PNM at the resonance frequency can be expressed as:

$$A = \frac{4\gamma_{\text{rad}}\gamma_{\text{int}}}{(\gamma_{\text{rad}}+\gamma_{\text{int}})^2}. \tag{1}$$

As qualitatively shown in Fig.1b, coupling to the incident light depends on the in-plane component of the nanofin's dipole moment. Since $h$ remains constant for all asymmetries, the nanofin's in-plane dipole moment is proportional to $\tan \alpha$, thus we define $AF$ as $\tan \alpha$ (Fig.2g) leading to $\gamma_{\text{rad}} \propto Q_{\text{rad}}^{-1} \propto AF^2 = \tan \alpha^2$. The intrinsic loss of our PNM can be extracted from Fig.2e at the CC condition, where $\gamma_{\text{int}} = \gamma_{\text{rad}}$ and thus $\gamma_{\text{int}} = \tan \alpha_{\text{CC}}^2$ with $\alpha_{\text{CC}} = 12°$. This gives rise to an analytical solution for the absorbance of the PNM:

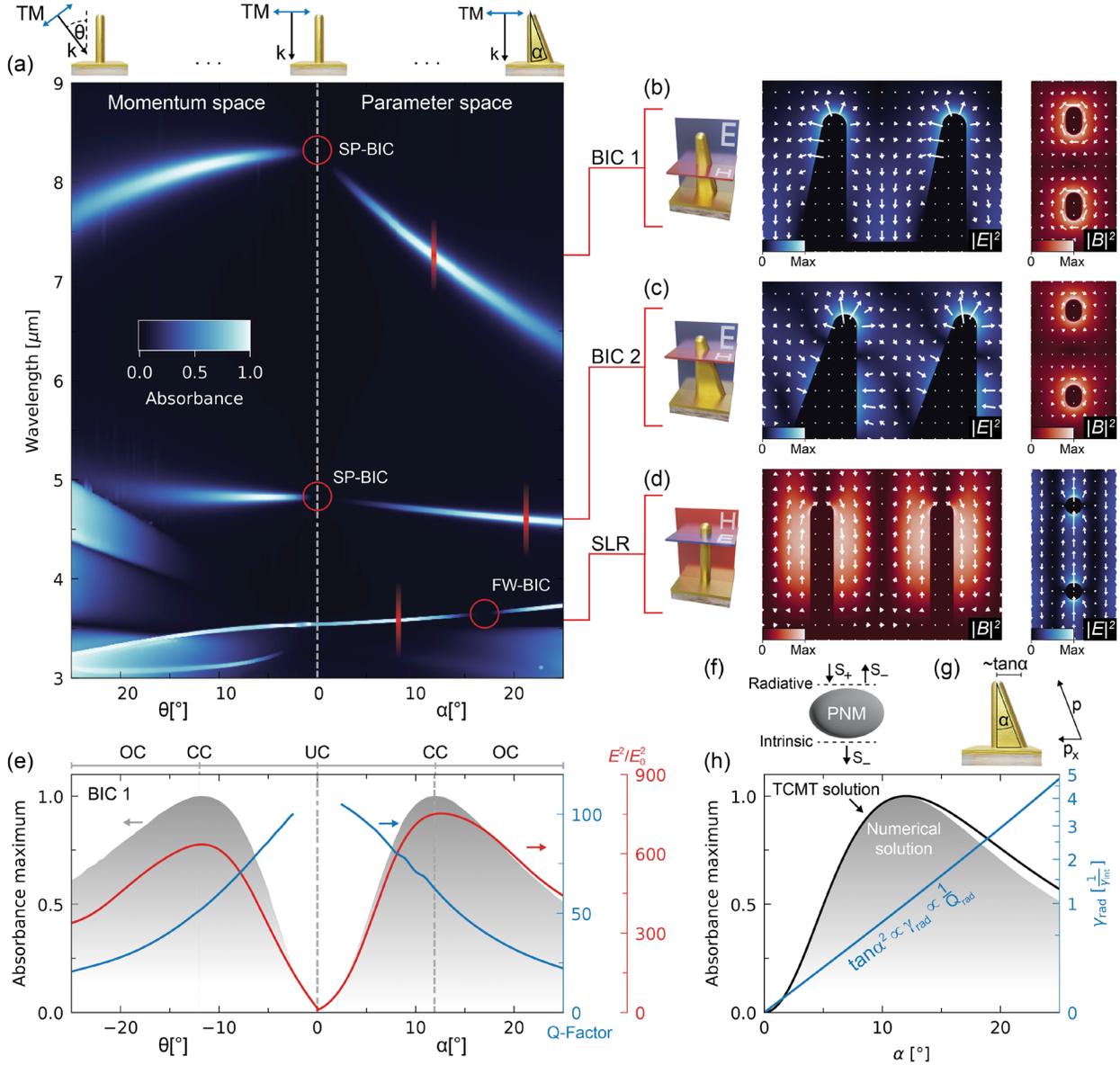

**Fig.2: Numerical analysis of symmetry-protected BICs in momentum and parameter space.** **(a)** MIR absorbance spectrum of a symmetric ($\alpha = 0°$) PNM in momentum space illuminated by light under an incident angle $\theta$ and of a PNM under symmetric excitation ($\theta = 0°$) in parameter space with a triangle angle $\alpha$. Three modes are visible in both spaces, which we attribute to 1st and 2nd order symmetry-protected BICs (BIC1 and BIC2) and a SLR (from long to short wavelengths, respectively). **(b-d)** Electric and magnetic field vectors (arrows) and intensities (in blue and red, respectively) of BIC1, BIC2, and the SLR. The fields were extracted according to the sketched cutting planes, and all correspond to the CC condition of each mode, marked as red lines in (a). **(e)** Maximal absorbance (grey), Q-factor (blue), and maximal field enhancement (red) of BIC1 in momentum ($\theta$) and parameter ($\alpha$) space. **(f)** PNM system illustrated as a single resonator coupled to one radiative and one intrinsic channel. **(g)** Nanofin sketch indicating the in-plane projection $p_x$ of the nanofin's dipole $p$. $p_x$ is proportional to the asymmetry

factor $\tan\alpha$. **(h)** The numerically obtained maximal absorbance (grey) is compared to the TCMT analytical model (black) assuming an inverse quadratic relation between the radiative Q-factor and an asymmetry factor $\tan\alpha$ (blue).

$$A = \frac{4\tan\alpha^2 \tan\alpha_{CC}^2}{(\tan\alpha^2 + \tan\alpha_{CC}^2)^2}. \qquad (2)$$

Without any further fitting, our analytical and numerical results match very well (Fig.2h). We find a nearly perfect agreement in the UC regime, while deviations remain below 10% in the OC regime, which might be attributed to a simplified asymmetry factor. The good agreement further confirms the relation of $Q_{\text{rad}}^{-1} \propto AF^2$ proofing the symmetry-protected BIC nature of our modes. Moreover, the asymmetry factor provides a pathway to precisely control the radiative loss of the PNM to tailor light-matter coupling from the UC to the OC regime.

**Fabrication and experimental verification**

The PNMs, each with areas of $160 \times 160\,\mu m^2$, were 3D laser-nanoprinted (two-photon polymerization) in liquid photoresist on a silica substrate (Fig.3a). After removing the unpolymerized resist, gold was sputtered from four different angles (Fig.3b). More details about the fabrication process are given in the Methods section. We fabricated PNMs with different $\alpha$ from the symmetric case of $\alpha = 0°$ to highly asymmetric with $\alpha = 26°$ in steps of 2°. The scanning electron microscopy images of our PNMs with $\alpha = 4, 12,$ and $22°$, associated with the UC, CC, and OC regimes, respectively, are given in Fig.3c. We find a good agreement between measured spectra (see Methods) of BIC1 and numerical results (Fig.3d and Fig.S5, Supplementary Note 3). The inset in Fig.3d shows CC around $\alpha = 12°$ with a maximum absorbance of 0.75 and a maximum Q-factor of 60. The small discrepancy between our experiments and simulations might result from an angular spread of the excitation source, imperfect linear polarization of light, finite PNM sizes, and fabrication deviations.

Our 3D nanoprint lithography approach allows us to continuously scale all size parameters as demonstrated in Fig.3e, where we sweep the BIC mode across a broad spectral range while keeping Q-factor and amplitude of the resonance nearly constant.

We tune critically-coupled PNMs ($\alpha = 12°$) with different scaling factors from 0.4 to 1.6 between $1.8 - 10\,\mu m$ (near- to MIR) limited only by the used detector (see Methods). Furthermore, scaling the height allows us to excite BIC modes up to the 4$^{th}$ order with even higher Q-factors of up to 84 (Supplementary Note 4, Fig.S6), which to our knowledge is, to date, the highest reported BIC order. We utilize the 2$^{nd}$ order BIC to perform refractive index sensing of a liquid (Fig.S7). We achieved a large experimental figure of merit (FOM) of 70, which is defined as $\text{FOM} = \frac{\Delta\lambda}{\lambda_{res}}$, where $\Delta\lambda$ and $\lambda_{res}$ represent the wavelength shift per refractive index unit and the resonance wavelength, respectively. Notably, for some fabricated PNMs, we can weakly couple light into the SLR (Fig.S8), however, the signal modulation is very low, which further indicates its instability regarding finite sample sizes and excitation angles.

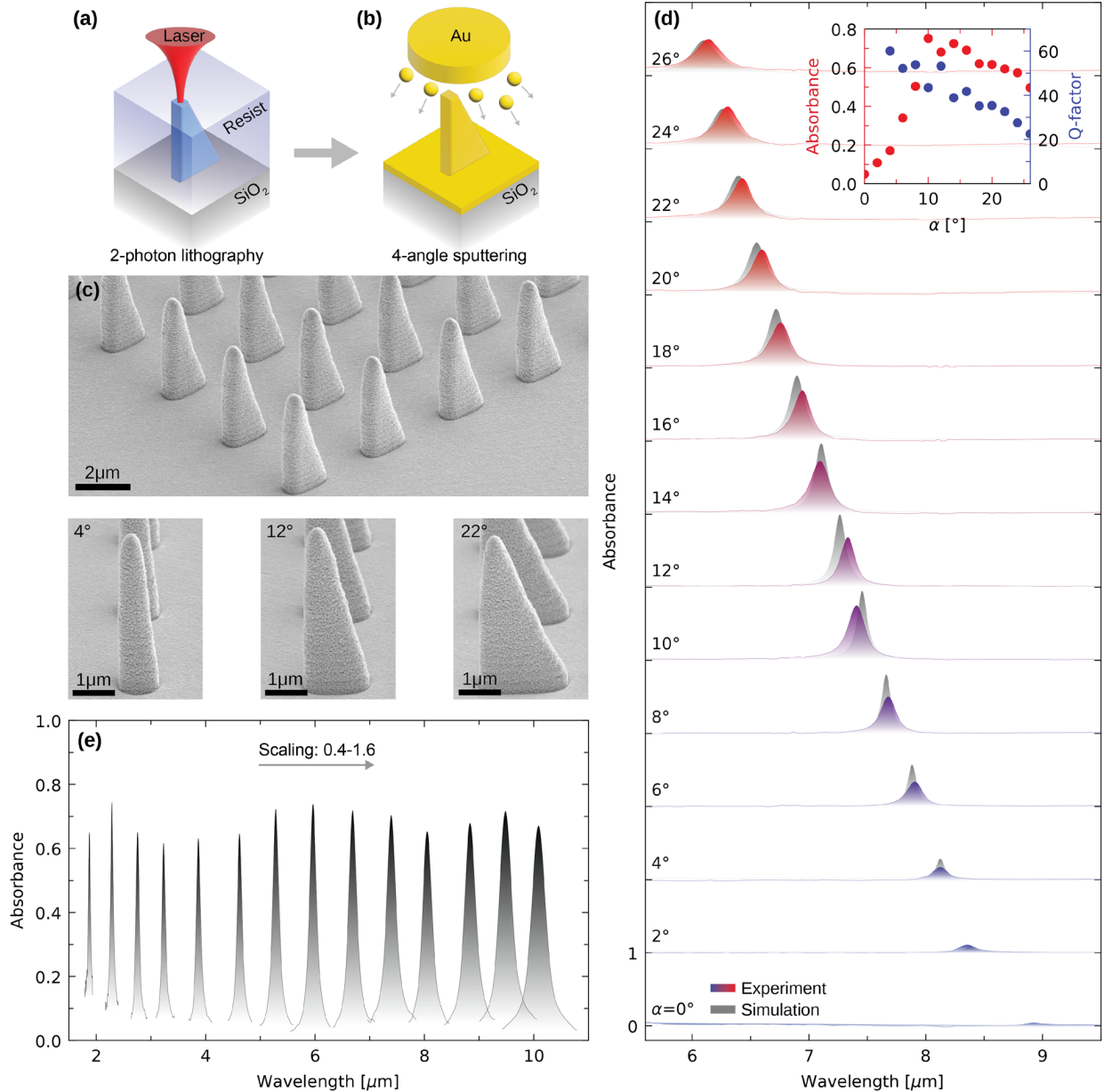

**Fig.3: Fabrication and experimental verification of higher-order BICs.** Schematic illustration of the two fabrication steps: **(a)** 3D laser nanoprinting in a "dip-in" configuration and **(b)** four-angle gold sputtering. **(c)** Scanning electron microscopy images of a PNM with $\Lambda = h = 3.5\mu m$, $d = 0.7\mu m$, and $\alpha = 12°$ in the top row, and side-view images for $\alpha = 4, 12,$ and $22°$ in the bottom row (the size parameters defining the PNM structure are given in Fig.S1). **(d)** Comparison of measured (colored) and simulated (grey) PNM spectra of BIC1 from symmetric ($\alpha = 0°$) to highly asymmetric ($\alpha = 26°$) structures. The inset shows corresponding amplitudes and Q-factors. **(e)** Ultrawide-scaled BIC1 spectra from the near-infrared ($1.8\mu m$) to the MIR ($10\mu m$) wavelength region.

**Molecular sensing based on a coupling-tailored PNM array**

Typically, surface-enhanced infrared absorption spectroscopy (SEIRAS) exploits the coupling of surface plasmon modes to weak MIR molecular vibrational absorption bands[28,38]. The molecules induce additional intrinsic loss to the resonator system, which dampens the plasmon mode around the absorption band leading to a highly enhanced signal[39]. It has been recognized that the field enhancement is the most critical feature for the sensitivity in SEIRAS[40,41]. To investigate the performance of our PNMs for molecular sensing, we designed and fabricated three coupling-tailored PNM arrays with $\alpha = 6, 12$, and $20°$ covering the UC, CC, and OC coupling regimes, respectively. To fully recover the spectral fingerprint of an analyte, we implemented the pixelated sensing approach[24] in each array through finely scaling the size of all PNM sensing units (Fig.4a).

To analyze our results, we adapted our TCMT model (Supplementary Note 2) by adding a second, non-radiative resonator to represent weak molecular vibrations (Fig.4b inset). Hence, this dark resonator acts as an additional intrinsic loss channel of the PNM. The adapted absorbance $A$ of the coupled PNM at the resonance frequency is given by:

$$A = \frac{4\gamma_{\text{rad}}(\gamma_{\text{int}} + \gamma_{\text{int2}})}{(\gamma_{\text{rad}} + \gamma_{\text{int}} + \gamma_{\text{int2}})^2} \quad (3)$$

with $\gamma_{\text{int2}} = \frac{\mu^2}{\gamma_{\text{analyte}}}$ as the additional intrinsic loss channel due to the analyte, where μ and $\gamma_{\text{analyte}}$ denote the coupling constant and the analyte's intrinsic loss, respectively. To study the effect of $\gamma_{\text{int2}}$, Fig.4b compares the calculated absorbance maximum as a function of $\alpha$ with (colored spectrum) and without (grey spectrum, same as in Fig.2e) the additional intrinsic loss channel based on equation (3). As an exemplary demonstration, we chose $\gamma_{\text{int2}} = 0.5\gamma_{\text{int}}$ in our calculations, which clearly shows the stretching of the curve, shifting the CC from $12°$ to $15°$.

Consequently, we show that the absorbance modulation depends strongly on the coupling regime of the PNMs. We chose polymethyl methacrylate (PMMA) as our analyte and investigate its prominent absorption line around $5.75\mu m$. Our numerical study (Figs.2c-2e) reveals a negative modulation for the UC regime and a positive modulation for the OC regime. Strikingly, we observe nearly no modulation around the CC condition.

This is in strong contrast to the intuitive assumption that the sensitivity for SEIRAS would be the highest at the CC position, where the field enhancement is the strongest. The envelopes of the simulated absorbance modulations in different coupling regimes were plotted in Fig.4f-h (lines), revealing a maximal modulation of around 15% (Fig.S9 for normalized modulation). For our measurements, we spin-coated a thin layer of PMMA on the PNM arrays (Methods). Our experimental results (dots in Fig.4f-h) follow the same trend in different coupling regimes as the simulations. As such, our demonstration explicitly proves that tailoring the coupling conditions of a high Q-factor metasurface is essential for high-performance molecular sensing.

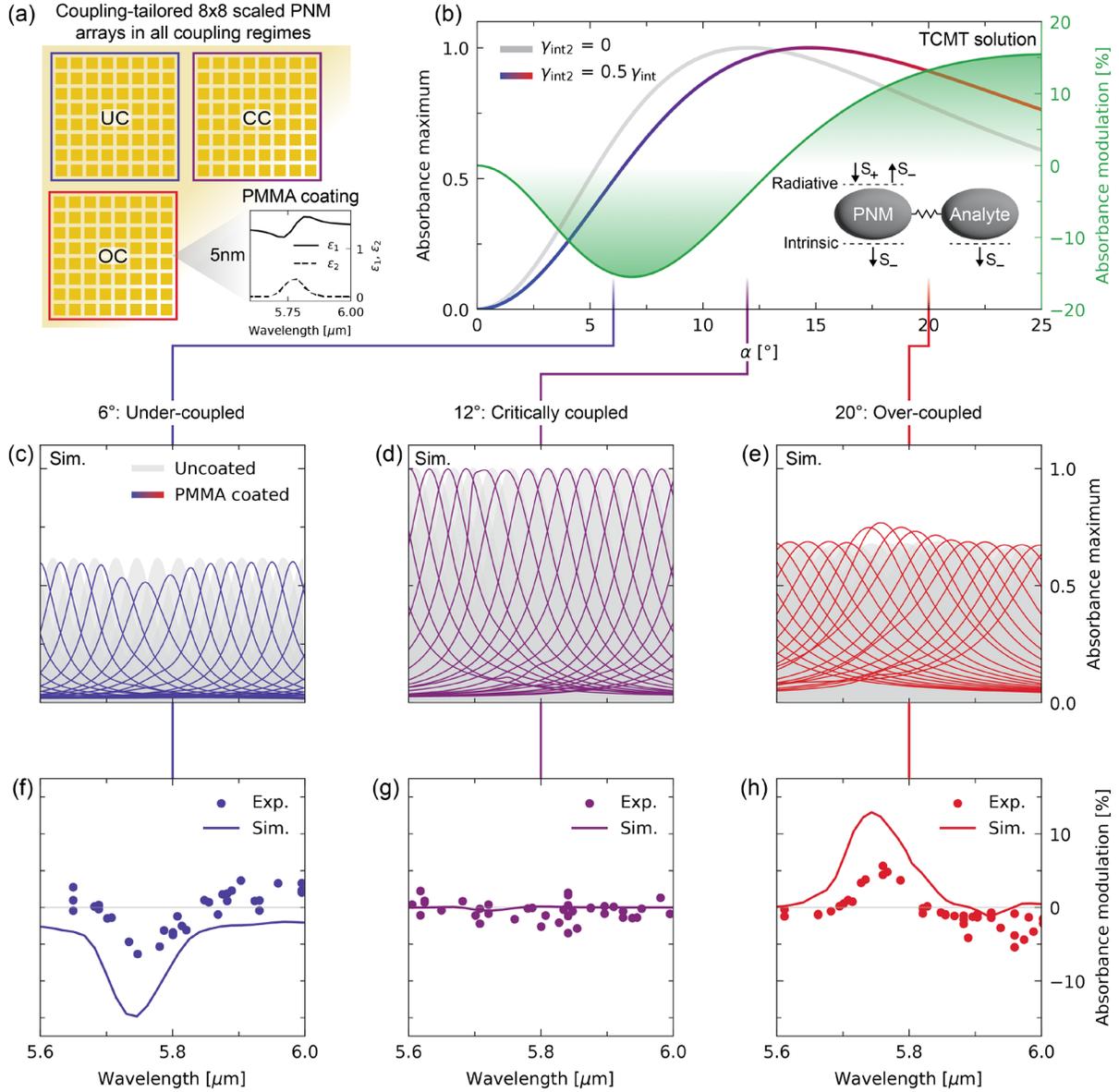

**Fig.4: Light-matter coupling-tailored molecular sensing. (a)** Illustration of three coupling-tailored PNM arrays, each with 8 by 8 size-scaled PNM units. **(b)** Analytical TCMT solution for the two-resonator system. The grey line denotes the $\alpha$-dependent maximal absorbance of the unperturbed bright mode, while the colored line shows the perturbed signal with additional loss $\gamma_{int2}$ due to the molecules. The absorbance modulation (the difference between the two lines) is depicted in green, revealing areas of negative, no, and positive modulation in the regions of UC, CC, and OC, respectively. **(c-e)** Uncoated (grey) and $5nm$ PMMA coated (colored) absorbance spectra for the UC, CC, and OC sensors, respectively. **(f-h)** The corresponding sensing results with both simulations (curves) and experiments (dots) consistently showing that the PMMA analyte can introduce negative, no, and positive modulations depending on the coupling regime.

**CONCLUSION**

We have designed a new PNM platform comprised of 3D triangular nanofins supporting out-of-plane symmetry-protected BICs in parameter space. Tuning the triangle angle $\alpha$ of the nanofins allows us to precisely tailor the radiative decay rate and couple to the metasurface in different regimes. We have shown that the coupling efficiency of the PNMs, represented by the metasurface absorbance, can be flexibly tuned from zero to unity. We fabricated our metasurface structures based on 3D laser nanoprinting in liquid resist. Our PNMs exhibit high Q-factor BIC modes up to the 4$^{th}$ order, scaled from 1.8 to 10 µm. In addition, we have designed and fabricated coupling-tailored PNM arrays for pixelated molecular sensing. Our results unveil a strong dependency of the sensitivity on the asymmetry parameter of the PNMs, which was echoed from our TCMT calculations. We show that the additional intrinsic loss imposed by the molecular analyte can introduce negative, positive, or even no modulations on the BIC resonance. As a result, we have demonstrated the importance of controlling light-matter interaction in a coupled metasurface system for enhancing the detection sensitivity of molecular sensing. We believe our demonstrated light-matter interaction-enhanced metasurface platform harnessing tunable high Q-factor plasmonic BICs paves the way for numerous applications for optical sensing[42], energy conversion[12], nonlinear photonics[14], surface-enhanced spectroscopy[5], quantum optics[43], and information technologies[44].

**METHODS**

All simulations were performed using the finite element solver *CST Studio Suite (Simulia, Providence, USA)* with adaptive mesh refinement and periodic boundary conditions. Since we assume an optically dense layer of gold, as discussed in Fig.S2, the PNM nanofins were simulated as solid gold triangles with no polymer core to reduce calculation time. We adapted the permittivity for gold and PMMA from Ref.[45] and Ref.[46], respectively, while we assume SiO$_2$ to have no dispersion with $n = 1.5$ and $k = 0$.

We printed the PNMs by using a *Photonic Professional GT 3D* (*Nanoscribe, Germany*) direct laser printer. A "dip-in" configuration was used with the substrate fixed upside down and liquid photoresist (*IP-Dip, Nanoscribe, Germany*) in between it and a 1.4NA, 63x oil immersion objective. A 780nm femtosecond pulsed laser polymerizes the resist via two-photon absorption processes. The focus position of the laser is controlled via a galvo mirror in the x-y-plane and a piezo stage along the z-axis. Afterwards, the unpolymerized resist is removed by placing the substrate in a PGMEA bath (15min). We further cleaned the samples in an IPA (15min) followed by a Novec 7000 (2min) bath. We used a *Von Ardenne LS 320 S magnetron sputterer* to sputter an adhesion layer of $\approx$ 5nm chrome followed by $\approx$ 40nm gold sputtered from four sides under an angle of $20°$.

The measurements in Fig.3e were performed with a *HYPERION II microscope (Bruker) combined with a VORTEX 80V FTIR (Bruker)*. We used a 15x mirror objective and a *LN-MCT detector* in reflection mode. For all other optical measurements, we used the spectral imaging MIR microscope Spero from *Daylight Solutions Inc., USA,* with a low 4x magnification objective ($NA = 0.1$) and a $2\text{mm}^2$ field of view. The system is equipped with four tunable quantum cascade lasers continuously covering the range between 5.6 and 10.5µm. To extract the spectral data from each metasurface array, we used an edge detection algorithm to identify the corresponding pixels. We determined the position, rotation, and size of single metasurfaces by fitting a grid mask over all selected pixels. Next, we averaged the spectral data of all pixels within the detected area to reduce noise. For the molecular sensing in Fig.4, we used a 1% solution (A1) of 495K PMMA in *Anisole* and spin-coated the film with 5000rpm (for 2min) followed by a 1min bake at 120°C.

**Acknowledgements:**

This work was funded by the Deutsche Forschungsgemeinschaft (DFG, German Research Foundation) under grant numbers EXC 2089/1 – 390776260 (Germany´s Excellence Strategy), the Bavarian program Solar Energies Go Hybrid (SolTech), and the Center for NanoScience (CeNS). A.T. acknowledges the Emmy Noether Program TI 1063/1. Y.K. acknowledges support from the Australian Research Council (projects DP200101168 and DP210101292). S. A. M. acknowledges the funding support from the Deutsche Forschungsgemeinschaft, the EPSRC (EP/W017075/1), and the Lee-Lucas Chair in Physics. H. R. acknowledges support from the Australian Research Council (project DE220101085).

**Author contributions**

A. A and H. R. conceived the idea; A. A. performed the numerical analysis, fabrication, and experiments; A. T. and J. W. contributed to the molecular sensing; T. W. contributed to the data processing; A. A., H. R., A. T., Y. K. and S. A. M. contributed to the data analysis; A. A. and H. R. wrote the paper with contributions from all authors.

# Supplementary Materials for

# Plasmonic Bound States in the Continuum to Tailor Light-Matter Coupling

Andreas Aigner[1,*], Andreas Tittl[1], Juan Wang[1], Thomas Weber[1], Yuri Kivshar[2], Stefan A. Maier[1,3,4,*], and Haoran Ren[3,*]

1) Chair in Hybrid Nanosystems, Nano-Institute Munich, Faculty of Physics, Ludwig-Maximilians-University Munich, Munich, 80539, Germany.
2) Nonlinear Physics Center, Research School of Physics Australian National University, Canberra, ACT 2601, Australia.
3) School of Physics and Astronomy, Monash University, Clayton, Victoria 3800, Australia.
4) Department of Physics, Imperial College London, London SW7 2AZ, United Kingdom.

*E-mails: andreas.aigner@physik.uni-muenchen.de; stefan.maier@monash.edu; haoran.ren@monash.edu

# SUPPLEMENTARY NOTES

**Note 1: Strong influence of the nanofin tip on the maximal surface field enhancement.** The maximal surface field enhancement given in Fig.2e is for a smooth spherical nanofin cap with $d = 700\text{nm}$. Reducing the surface of the cap in any way will drastically enhance the maximal surface field enhancement. To illustrate this, we simulated PNMs with elongated caps (top sphere scaled in z-direction by a factor of 2), with $d = 500\text{nm}$, and with conical tips, yielding maximal surface field enhancements of 1380, 1690, and 182000, respectively. Thus, our maximal surface field enhancement values as well as most values claimed by others, especially in the IR, need to be taken with a grain of salt.

**Note 2: Derivation of the temporal coupled-mode theory (TCMT) model.** In the following, we adapt the generally formulated TCMT of optical resonators[1]. The dynamic equations of a coupled resonator system can be written as:

$$\frac{d\vec{a}}{dt} = (i\Omega\text{-}\Gamma)\vec{a} + K^{\dagger}|s_{+}\rangle \qquad (S1)$$

$$|s_{-}\rangle = C|s_{+}\rangle + D\vec{a} \qquad (S2)$$

with $\vec{a}$ as the amplitude vector of the modes and $\Omega$ containing the eigenfrequencies and mode-to-mode coupling. $\Gamma$ describes the energy decay, $K$ the port-to-mode coupling, $C$ the port-to-port coupling, and $D$ the mode-to-port coupling while $|s_{+}\rangle$ and $|s_{-}\rangle$ represent incoming and outgoing waves.

For our system, we consider two modes: the plasmon (mode1) and the vibrational absorption (mode2) of the analyte. While mode1 is coupled to one radiation and one intrinsic channel (bright), mode2 is only connected to an intrinsic channel (dark). Considering intrinsic channels as ports with only outgoing energy results in

$$\vec{a} = \begin{pmatrix} a_1 \\ a_2 \end{pmatrix}, \Omega = \begin{pmatrix} \omega_1 & \mu \\ \mu & \omega_2 \end{pmatrix}, \Gamma = \begin{pmatrix} \gamma_{11} & 0 \\ 0 & \gamma_{22} \end{pmatrix}, D = \begin{pmatrix} d_{11} & 0 \\ d_{21} & 0 \\ 0 & d_{32} \end{pmatrix}, \text{ and } |s_+\rangle = \begin{pmatrix} s_{1+} \\ 0 \\ 0 \end{pmatrix}.$$

Energy conversion and time reversal symmetry yields the following three relations[1] $D^{\dagger}D = 2\Gamma$, $K = D$, $CD^{*} = \text{-}D$, which further lead to

$$\Gamma = \frac{1}{2}D^{\dagger}D = \frac{1}{2}\begin{pmatrix} d_{11}^2 + d_{21}^2 & 0 \\ 0 & d_{32}^2 \end{pmatrix}, \quad K^{\dagger} = \begin{pmatrix} k_{11} & 0 \\ k_{21} & 0 \\ 0 & k_{32} \end{pmatrix} = \begin{pmatrix} d_{11} & 0 \\ d_{21} & 0 \\ 0 & d_{32} \end{pmatrix}, \quad C = -\begin{pmatrix} 1 & 0 & 0 \\ 0 & 1 & 0 \\ 0 & 0 & 1 \end{pmatrix},$$

and $|s_-\rangle = \begin{pmatrix} s_{1+} \\ 0 \\ 0 \end{pmatrix} + D\vec{a}$.

The energy decay for mode1 $\gamma_{11}$ consists of two parts: $\frac{1}{2}d_{11}^2$ and $\frac{1}{2}d_{21}^2$ representing radiation and intrinsic loss, respectively, so we define $\gamma_{rad1} = \frac{1}{2}d_{11}^2$ and $\gamma_{int1} = \frac{1}{2}d_{21}^2$ (for mode2: $\gamma_{22} = \gamma_{int2} = \frac{1}{2}d_{32}^2$), which yields the dynamic equations

$$i\omega a_1 = \left(i\omega_1 - \frac{1}{2}(d_{11}^2 + d_{21}^2)\right)a_1 + i\mu a_2 + d_{11}s_{1+} \quad (S3)$$

$$i\omega a_2 = \left(i\omega_2 - \frac{1}{2}d_{32}^2\right)a_2 + i\mu a_1. \quad (S4)$$

Substituting $a_2$ in (S3) using (S4) and setting $\omega = \omega_1 = \omega_2$ results in

$$0 = -\frac{1}{2}\left(d_{11}^2 + d_{21}^2 + \frac{2\mu^2}{d_{32}^2}\right) + d_{11}s_{1+} = -\left(\gamma_{rad1} + \gamma_{int1} + \frac{\mu^2}{\gamma_{int2}}\right) + d_{11}s_{1+} \text{ and}$$

$$|s_+\rangle = \begin{pmatrix} s_{1+} + d_{11}a_1 \\ d_{21}a_1 \\ d_{32}a_2 \end{pmatrix} = \begin{pmatrix} s_{1+} + \sqrt{2\gamma_{rad1}}a_1 \\ \sqrt{2\gamma_{int1}}a_1 \\ \sqrt{2\gamma_{int2}}a_2 \end{pmatrix}.$$

Now we can use equation (S1) and (S2) to find a solution for the reflectance R and thus a solution for the absorbance A:

$$A = 1-R = 1-\left|\frac{s_-}{s_+}\right|^2 = 1-\left|\frac{(-s_{1+}-d_{11}a_1)}{s_{1+}}\right|^2 = \frac{4\gamma_{rad1}\left(\gamma_{int1} + \frac{\mu^2}{\gamma_{int2}}\right)}{\left(\gamma_{rad1} + \gamma_{int1} + \frac{\mu^2}{\gamma_{int2}}\right)^2}. \quad (S5)$$

For a single resonator without analyte, the absorbance simplifies to

$$A = \frac{4\gamma_{rad1}\gamma_{int1}}{(\gamma_{rad1}+\gamma_{int1})^2}. \quad (S6)$$

**Note 3: Adapting the simulation parameters to the fabricated structures.** When comparing the cylindrical shapes of the simulated PNM with the fabricated ones, two differences are apparent which need to be accounted for. The first one is the not perfect cylindrical shape, which is a result of the laser writing process where each nanofin layer is printed by multiple lines of equal length, which results in a slightly (in-plane) rectangular shape. The second deviation comes from an α-dependent exposure time to the laser. Since all nanofins are printed using the same parameters for power and speed, nanofins with larger α receive more laser intensity (larger in-plane rectangles) and thus their diameter and top width is increased compared to those with a smaller α. We compensate these effects by measuring the diameter $d$ and the top width in x-direction $d_x$ for 0° and 26° samples and interpolate the values in between with a linear function and smoothing the triangle edges with a $d_x$-dependent radius $r$.

The following table lists the used parameters while Fig.S5 sketches the simulated model:

|  | Value at 0° | Value at 26° | Linear interpolation |
|---|---|---|---|
| d [nm] | 635 | 780 | $635 + 5.58 \cdot \alpha$ |
| $d_x$ [nm] | 410 | 600 | $410 + 7.31 \cdot \alpha$ |
| r [nm] | $\frac{1}{2} d_x - 20 = 185$ | $\frac{1}{2} d_x - 20 = 280$ | $185 + 3.65 \cdot \alpha$ |

**Note 4: Higher-order BICs.** It is known that the metasurface unit cell must be subwavelength to prevent higher-order diffraction effects, so the resonance wavelength $\lambda_{res}$ should satisfy: $\lambda_{res} > n_s \cdot \Lambda$, where $n_s$ and $\Lambda$ are the substrate refractive index and pitch, respectively. It turned out to be extremely challenging for conventional in-plane symmetry-broken geometries to access to higher-order BICs, as finding resonators supporting in-plane multipole moments in a subwavelength scale is difficult. To date, only lower-order BIC modes (1st and 2nd orders) have been realized from high refractive index resonators[2]. On the other hand, for our out-of-plane symmetry-broken PNM, the dipole moments do not lie within the metasurface plane. We show that 3rd and 4th order BICs can be achieved for PNMs with $\Lambda = 6\mu m$ and aspect ratios of 1.33 and 2, respectively (Fig.S6a,b). All the BIC modes feature an absorbance above 0.5 and Q-factors of up to 84. The simultaneous access to multiple BIC modes based on our PNM platform opens the possibility to develop multifunctional photonic devices including multicomponent sensors[3].

## Supplementary Figures

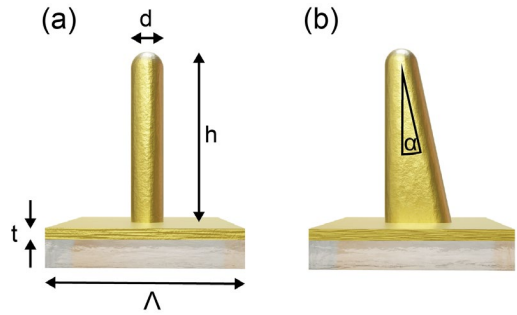

**Fig.S1:** Sketch of a single PNM unit cell, defined by the pitch $\Lambda$, the height h, the diameter d, the triangle angle $\alpha$, and the gold film thickness t.

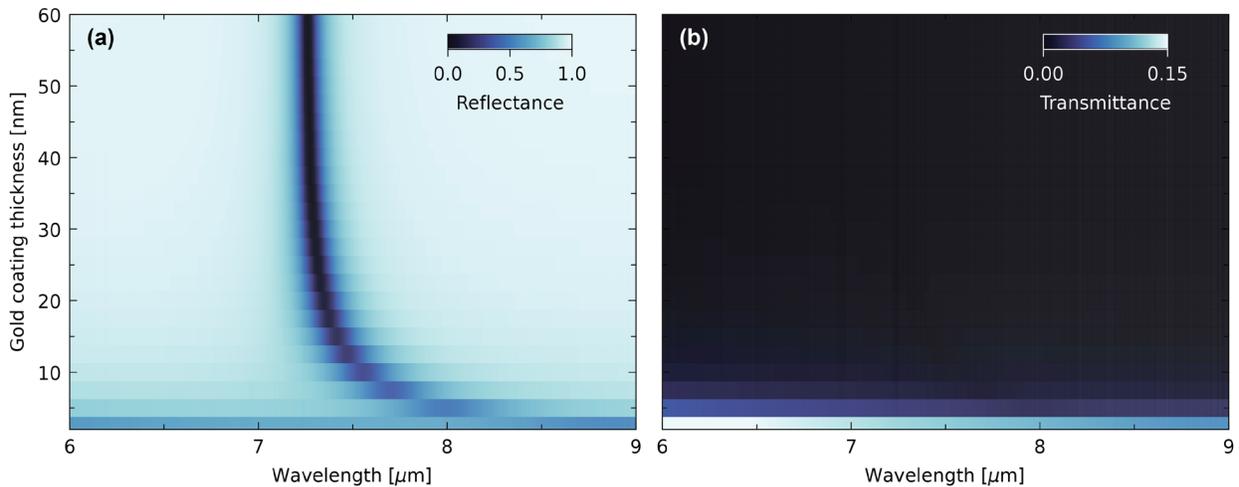

**Fig.S2: (a)** Reflectance and **(b)** transmittance spectrum around the BIC mode (p = h = 3.5μm, d = 0.7μm, $\alpha$ = 12°) for different gold coating thicknesses in 2D color plots. The transmittance quickly drops to zero for thicknesses above 10nm while the BIC lineshape, visible in the reflectance spectrum, becomes narrower and shifts for increasing thicknesses up to 30nm. For larger values the mode remains constant so we can assume an optically dense layer of gold for thicknesses larger than 30nm.

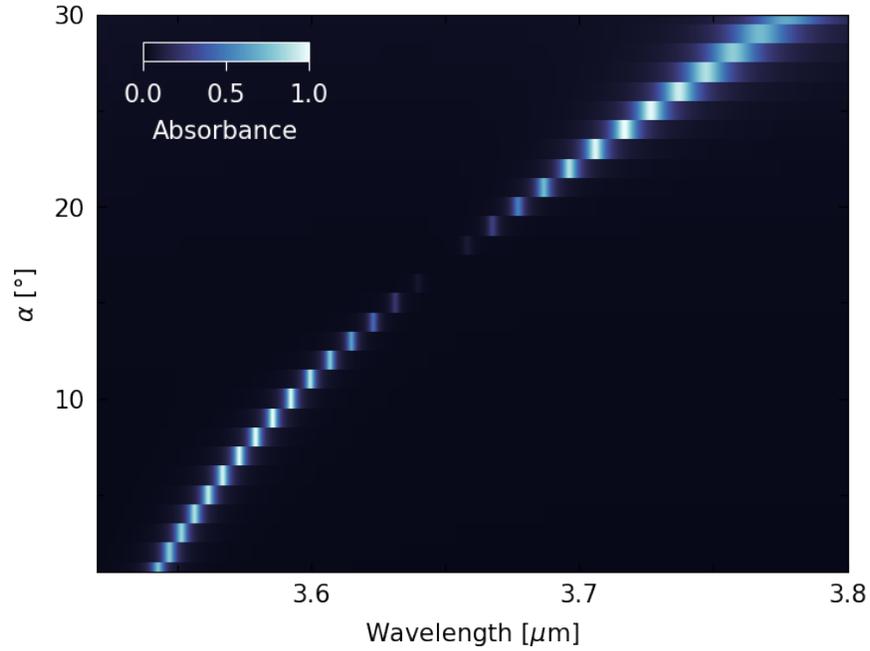

**Fig.S3:** Close-up of the SLR absorbance color plot of Fig.2a for different asymmetries in parameter space featuring a BIC around $\alpha = 18°$.

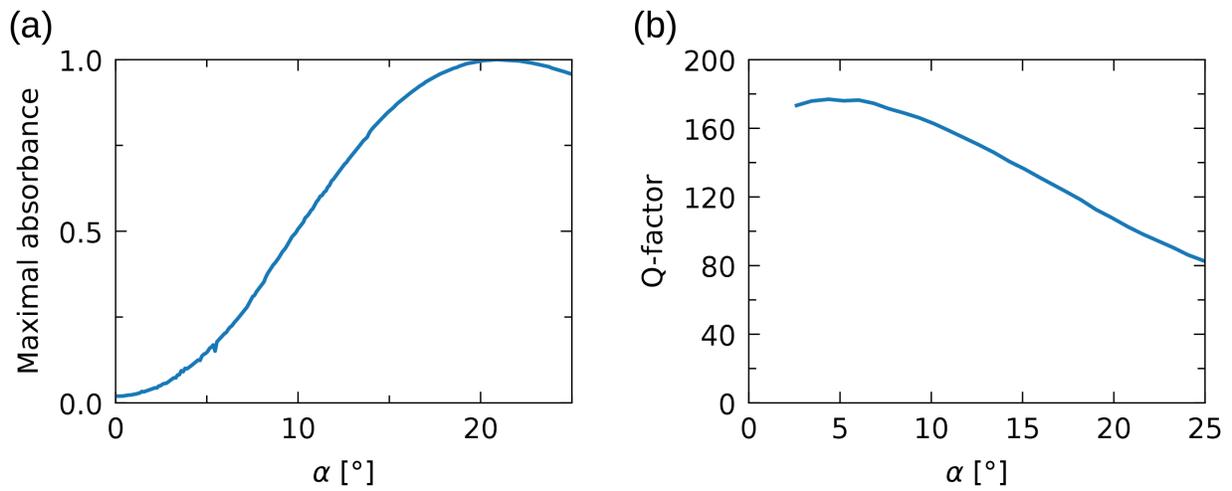

**Fig.S4: (a)** Maximal absorbance and **(b)** Q-factors of BIC2 for $\alpha = 0\text{-}25°$ corresponding to Fig.2a with $p = h = 3.5\mu m$ and $d = 0.7\mu m$.

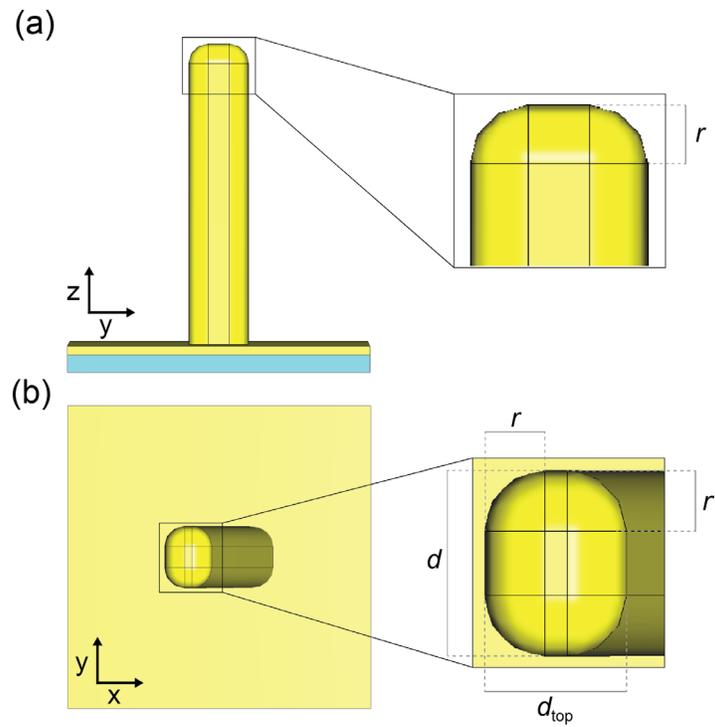

**Fig.S5:** Sketch of the PNM unit cell used for simulations in Fig.3b with the adaptations according to the fabricated structures. **(a)** Side view (y-z-plane) and **(b)** top view (x-y-plane) with the widths $d$ and $d_x$ and the smoothing radius $r$.

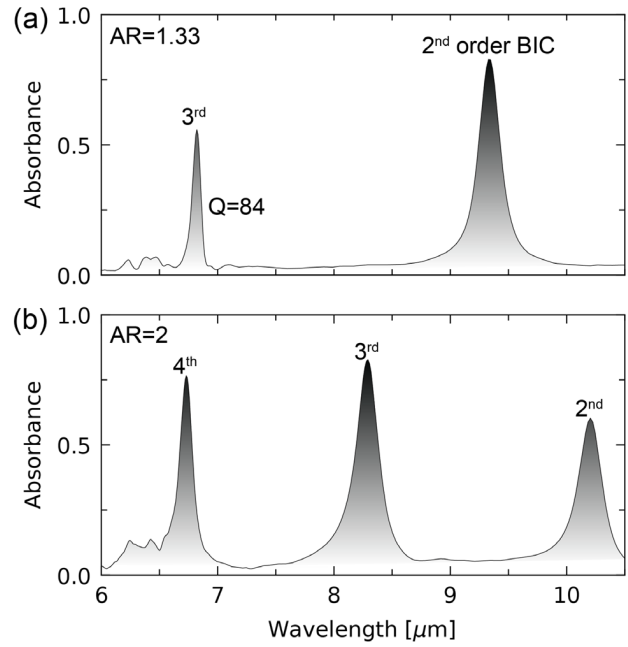

**Fig.S6:** Absorption spectra for PNMs with $\Lambda = 6\mu m$ and aspect ratios $AR$ between height $h$ and pitch $\Lambda$ of **(a)** 1.33 and **(b)** 2, revealing BIC modes up to the 3$^{rd}$ and 4$^{th}$ order, respectively.

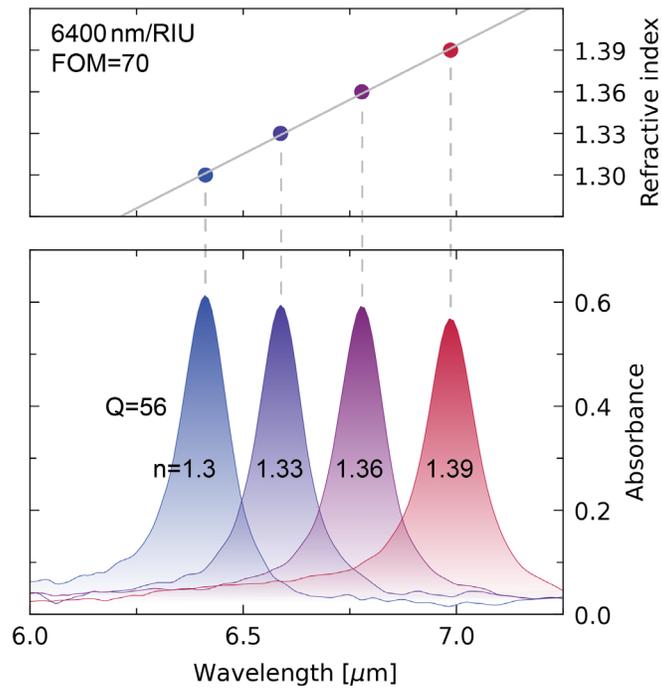

**Fig.S7:** PNM immersed in refractive index liquids with n = 1.3-1.39 in four steps utilizing the 2$^{nd}$ order BIC resonance, yielding 6400nm/RIU and a FOM of 70. This value is competitive with current sensor designs[4,5].

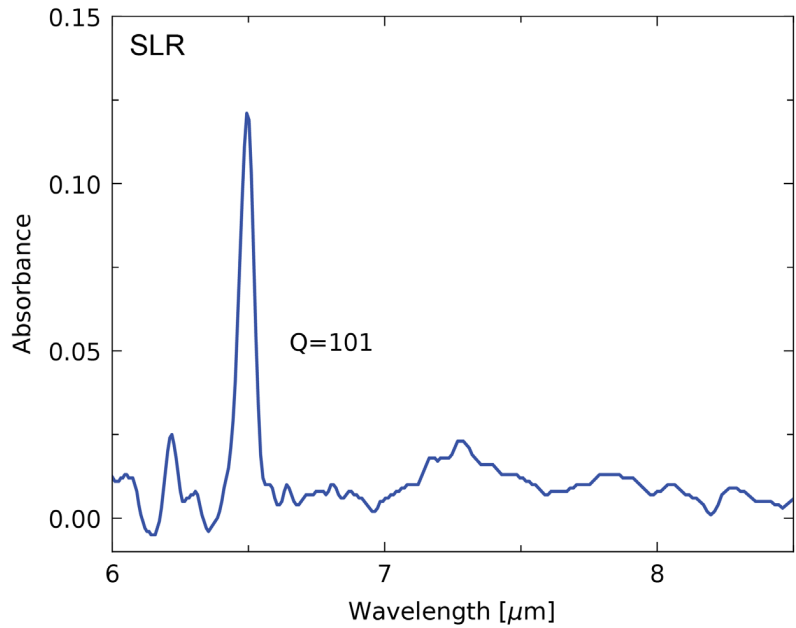

**Fig.S8:** SLR of a fabricated PNM with p = 6μm, h = 7μm, d = 0.9μm, and α = 0°. This mode only appeared this pronounced in a handful of structures, and for most PNMs the SLR was damped almost completely.

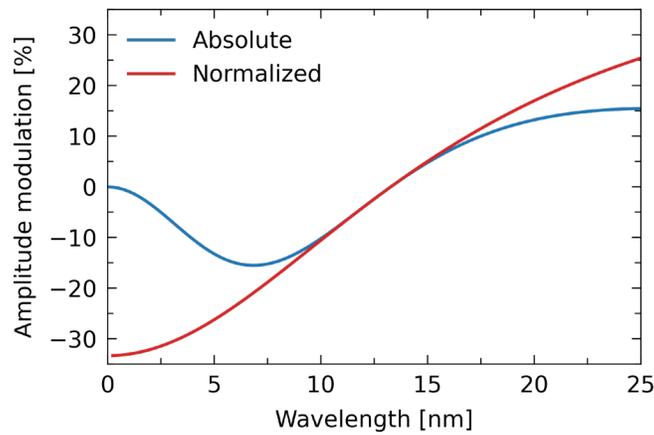

**Fig.S9:** Absolute ($A_{2res}-A_{1res}$) and normalized ($\frac{A_{2res}-A_{1res}}{A_{1res}}$) change in absorbance of Fig.4b with $A_{1res}$ as the absorbance of the one-resonator system and $A_{2res}$ of the two-resonator system.